\documentclass[preprint,aps,prl,showpacs,amsfonts,amssymb,superscriptaddress]{revtex4}
\usepackage{graphicx}
\usepackage{stmaryrd}

\begin{document}
\title{Charge induced coherence between intersubband plasmons in a quantum structure}
\author{A. Delteil}
\affiliation{Univ Paris Diderot, Sorbonne Paris Cit\'e, Laboratoire Mat\'eriaux et Ph\'enom\`enes Quantiques, UMR7162, 75013 Paris (France)}
\author{A. Vasanelli}
\email{angela.vasanelli@univ-paris-diderot.fr}
\affiliation{Univ Paris Diderot, Sorbonne Paris Cit\'e, Laboratoire Mat\'eriaux et Ph\'enom\`enes Quantiques, UMR7162, 75013 Paris (France)}
\author{Y. Todorov}
\affiliation{Univ Paris Diderot, Sorbonne Paris Cit\'e, Laboratoire Mat\'eriaux et Ph\'enom\`enes Quantiques, UMR7162, 75013 Paris (France)}
\author{C. Feuillet Palma}
\affiliation{Univ Paris Diderot, Sorbonne Paris Cit\'e, Laboratoire Mat\'eriaux et Ph\'enom\`enes Quantiques, UMR7162, 75013 Paris (France)}
\author{M. Renaudat St-Jean}
\affiliation{Univ Paris Diderot, Sorbonne Paris Cit\'e, Laboratoire Mat\'eriaux et Ph\'enom\`enes Quantiques, UMR7162, 75013 Paris (France)}
\author{G. Beaudoin}
\affiliation{Laboratoire de Photonique et Nanostructures, CNRS, 91460 Marcoussis, France}
\author{I. Sagnes}
\affiliation{Laboratoire de Photonique et Nanostructures, CNRS, 91460 Marcoussis, France}
\author{C. Sirtori}
\email{carlo.sirtori@univ-paris-diderot.fr}
\affiliation{Univ Paris Diderot, Sorbonne Paris Cit\'e, Laboratoire Mat\'eriaux et Ph\'enom\`enes Quantiques, UMR7162, 75013 Paris (France)}

\date{\today}

\begin{abstract}
In this work we investigate a low dimensional semiconductor system, in which the light-matter interaction is enhanced by the cooperative behavior of a large number of dipolar oscillators, at different frequencies, mutually phase locked by Coulomb interaction. We experimentally and theoretically demonstrate that, owing to this phenomenon, the optical response of a semiconductor quantum well with several occupied subbands is a single sharp resonance, associated to the excitation of a {\emph{bright multisubband plasmon}}. This effect illustrates how the whole oscillator strength of a two-dimensional system can be concentrated into a single resonance independently from the shape of the confining potential. When this cooperative excitation is tuned in resonance with a cavity mode, their coupling strength can be increased monotonously with the electronic density, allowing the achievement of the ultra-strong coupling regime up to room temperature.  
\end{abstract}

\maketitle
Elementary excitations in dense media are renormalized by their mutual interaction, giving rise to quasi-particles with new collective properties. This quantum interaction occurring at the particle scale manifests itself by powerful phenomena with macroscopic properties that can be easily measured. This is a common situation in condensed matter. Well known examples are the divergence of the conductivity in superconductor metals~\cite{onnes} associated to Cooper pair formation~\cite{bardeen}, the observation of absorption resonances induced by the collective excitation of electrons in plasma~\cite{raether} and the enhanced superradiant emission from a collection of particles~\cite{dicke}. For all these phenomena the quantum systems are in a coherent superposition of states that can lead to interference effects, observable in transport or optical spectroscopy.
In this work we focus on the optical response of a low dimensional semiconductor system in which a large number of electronic dipoles, although oscillating at very different frequencies, can be phase-locked by Coulomb interaction, to produce a giant optical polarization. Our system is based on a highly doped semiconductor quantum well, with several occupied subbands. As a consequence of the charge induced coherence, the electronic intersubband transitions associated to the well become transparent and collapse into a single sharp resonance, as we demonstrate both experimentally and theoretically. The optical properties of the system are thus no longer tied to the intrinsic energy levels of the structure, but derive from interference phenomena occurring between the electronic states. When this collective excitation, with optical transitions in the mid infrared, is tuned in resonance with a cavity mode, the system enters the ultra-strong coupling regime, up to room temperature. Our model, applied to a parabolic potential, verifies the Kohn theorem and illustrates how the dynamic energy renormalization (depolarization) is exactly compensated by the static (Hartree) charge effects.

\begin{figure}[ht]
\centering
\includegraphics[width=0.8\columnwidth]{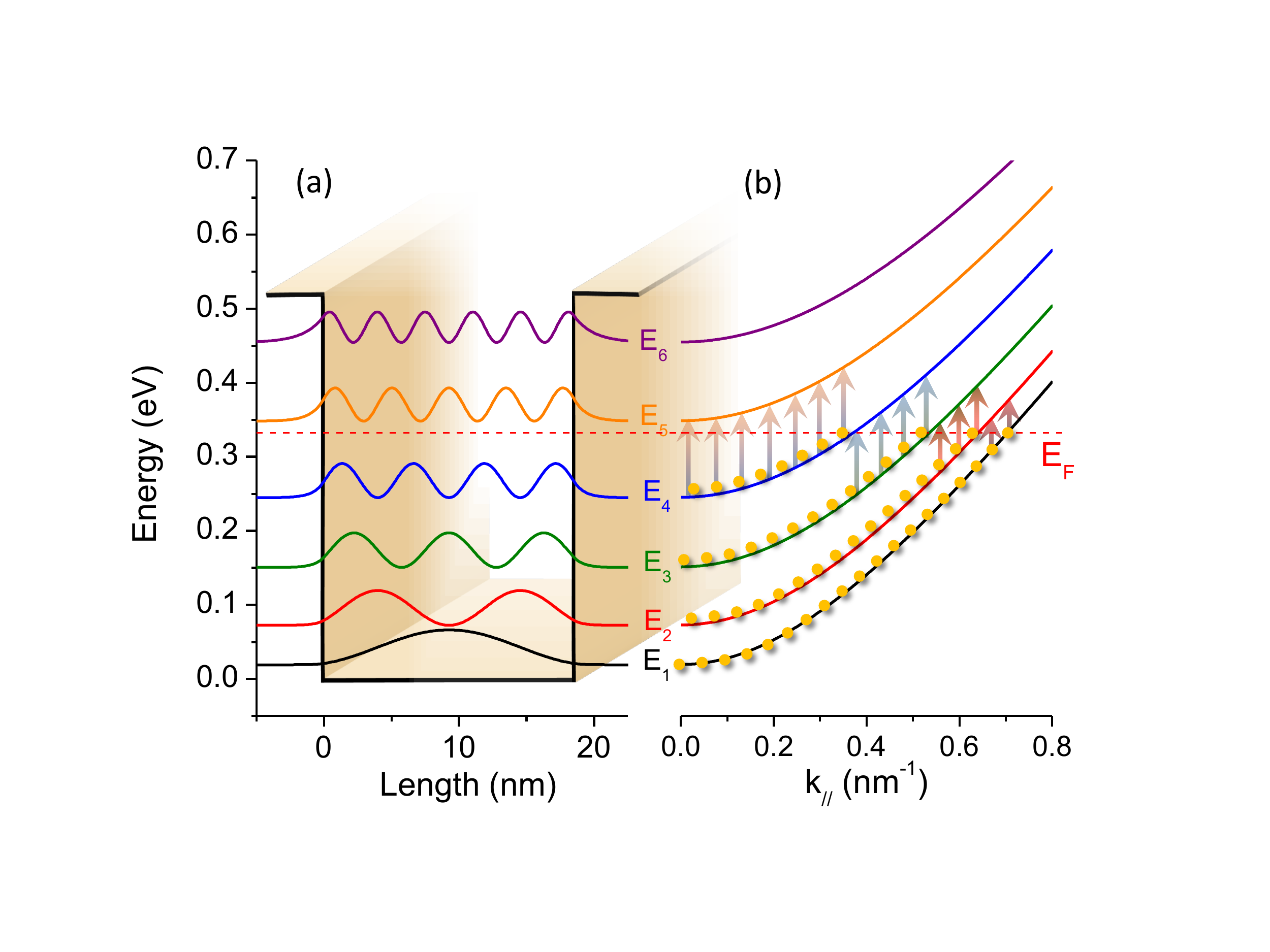}
\caption{(a) Band structure and energy levels of an 18.5 nm GaInAs/AlInAs quantum well. (b) Electronic dispersion of the subbands. The red dashed line indicates the Fermi energy at 0 K in our sample. The arrows indicate the main optical transitions that can take place in the structure.}
\label{fig:qw}
\end{figure}

The effect of the charge induced coherence has been investigated by analyzing the optical response of a single GaInAs/AlInAs semiconductor quantum well 18.5 nm thick, grown on an InP substrate by Metal Organic Chemical Vapor Deposition. Figure 1a presents the band structure and energy levels of the quantum well in the direction of the confinement $z$. In figure~\ref{fig:qw}b we show the electronic dispersion as a function of the in-plane momentum, $k_{//}$, calculated by using a three band $\vec{k} \cdot \vec{p}$ model~\cite{sirtori}. The structure is doped in the well with an electronic density of $2.2 \times 10^{13}$~cm$^{-2}$, resulting in a Fermi energy at 0 K above the 4th level ($E_F=331$~meV, indicated by a dashed line in Fig.~\ref{fig:qw}). The single particle transitions at $k_{//} = 0$ correspond to the energies:  $E_{12}=53.3$~meV, $E_{23}=78.1$~meV, $E_{34}=94.2$~meV, $E_{45}=103.4$~meV, $E_{56}=106.2$~meV. These transition energies are calculated in the Hartree approximation and are very similar to those that one would obtain in a square quantum well, as the entire well is doped~\cite{nota1}. 
Using a single particle description and a Lorentzian line broadening for each allowed transition, the absorption spectrum would appear as the blue line in figure 2. This very broad feature would be the result of a collection of independent and incoherent oscillators at the frequencies of the different transitions~\cite{nota2}, schematized by vertical arrows in fig.~\ref{fig:qw}b.

\begin{figure}[ht]
\centering
\includegraphics[width=0.9\columnwidth]{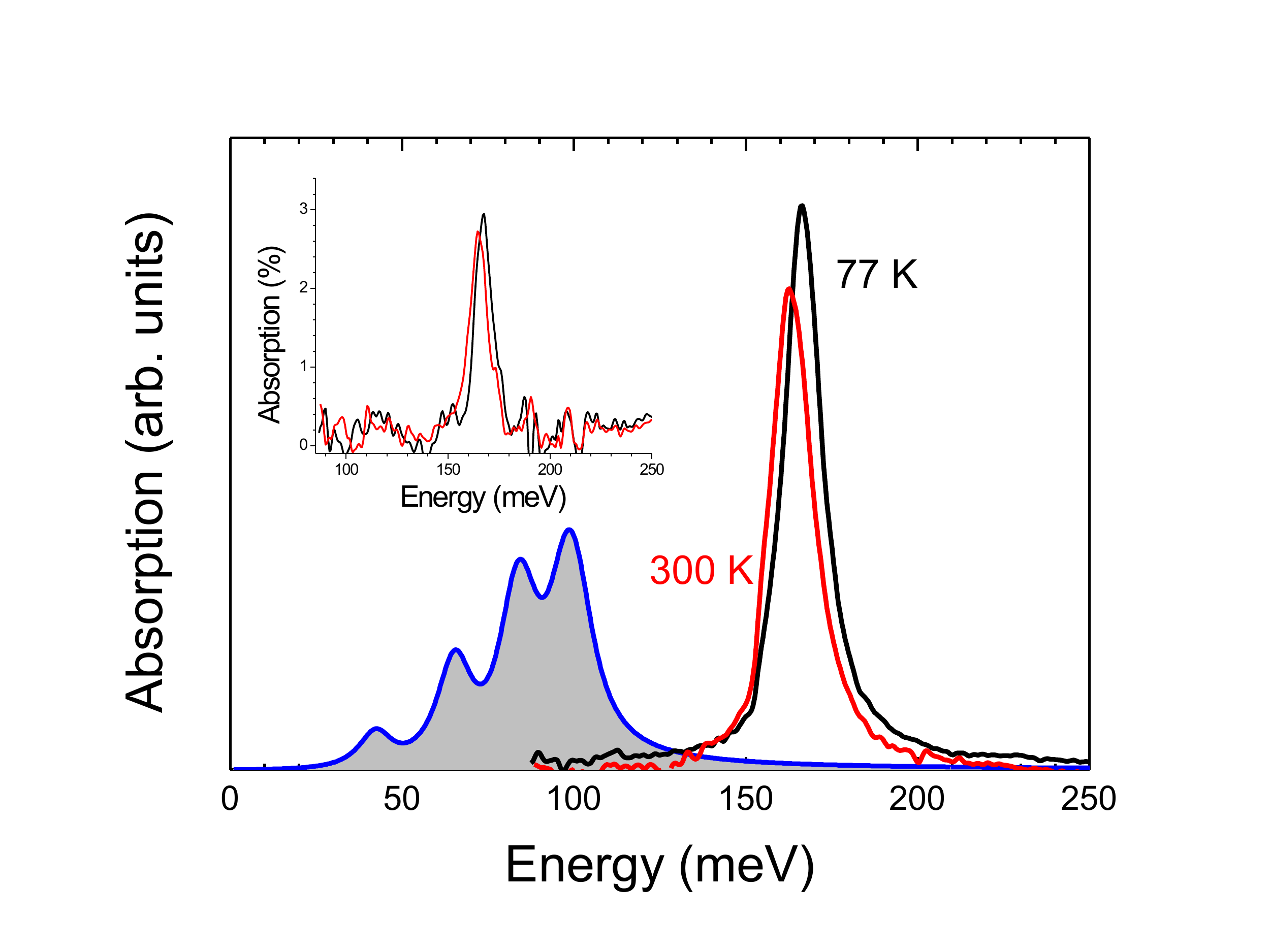}
\caption{Absorption spectra of the 18.5 nm GaInAs/AlInAs doped quantum well measured at 77 K (black line) and 300 K (red line). The inset presents the 77K (black line) and 300K (red line) spectra measured on the same sample at Brewster angle. The blue line in the main panel represents the simulated absorption spectrum, resulting from a single particle description and Lorentzian line broadening of the allowed transitions. }
\label{fig:spectra}
\end{figure}

The main panel of Fig.~\ref{fig:spectra} presents the absorption spectra measured at 77 K (black line) and 300 K (red line) by using a multipass geometry on a sample with $45^ \circ$ polished facets. Both spectra are characterized by a single sharp resonance. In order to precisely measure the linewidth we have to avoid eventual spatial inhomogeneities of the sample and saturation effects due to the strong absorption. We have thus measured transmission spectra also at Brewster angle, as presented in the inset of figure~\ref{fig:spectra}. In this configuration the full widths at half maximum are $\Delta E/E = 5.1\%$ at 77 K and $\Delta E/E = 5.7\%$ at 300K, respectively. In the spectra there is no trace of the transitions $E_{34}$ and $E_{45}$, which would occur within the range of our detection. The data illustrate that our system has become transparent for all the transition energies between adjacent levels~\cite{li}, expected from a single particle picture. Moreover, other mechanisms that should increase the broadening, such as the non-parabolicity and the high concentration of ionized impurities in the well, do not affect the transition linewidth, which stays comparable with that of intersubband transitions measured in modulation doped samples. All these features are indicative of a many-body optical resonance, arising from the collective oscillations of electrons coupled by Coulomb interaction ~\cite{metzner, zaluzny, luin}. In our system this interaction lumps together all the transitions of the well into a single collective bright excitation of the 2D electron gas (2DEG) resulting in a spectacular renormalization of the optical response that has no intuitive relation with the transitions between electronic energy levels. 

To explain the physical origin of the collapse of the single particle transitions into a single bright excitation, we describe the light-matter interaction in the dipole gauge Hamiltonian~\cite{todorov_PRB2012}, which contains the effects of the dipole-dipole interactions and the coupling of the electronic polarization with light. The diagonalisation of the Hamiltonian can be described in a three steps procedure. The first two concern the matter polarization and only in the third phase light-matter interaction is treated. The first operation is to express the matter excitation in terms of the intersubband plasmons \cite{ando, wendler, allen, chen, nikonov, graf}. In this first step the transition energies $E_{\alpha}$ (where $\alpha$ labels the intersubband transitions) are renormalized by their plasma energy $E_{P_{\alpha}}=\hbar e \sqrt{\frac{f_{\alpha} \Delta N_\alpha }{m^* \epsilon_0 \epsilon_{st} L_{eff}}}$ as $\widetilde{E}_{\alpha}=\sqrt{E_{\alpha}^2+ E_{P_{\alpha}}^2}$. Here $\Delta N_\alpha$ is the difference between the electronic densities of the initial and final subbands involved in the transition, $f_\alpha$ is its oscillator strength, $\epsilon_{st}$ is the static background dielectric constant, $e$ the electron charge, $m^*$ the constant effective mass and $L_{eff}$ is the effective quantum well thickness renormalized by the Coulomb interaction~\cite{ando}. 
At this stage the matter part of the Hamiltonian is written as:
\begin{equation}
\label{Hmatter}
H_{\mathrm{matter}} = \sum _\alpha {\widetilde{E}_{\alpha} p_{\alpha}^\dagger p_{\alpha}} + \frac{\hbar}{2} \sum_{\alpha \neq \beta}{\Xi_{\alpha,\beta} \left( p_{\alpha}^\dagger + p_{\alpha}\right) \left( p_{\beta}^\dagger + p_{\beta}\right)} 
\end{equation}
where the operator $p_{\alpha}^\dagger$ (resp. $p_\alpha$) describes the creation (resp. annihilation) of an optically active plasmonic excitation associated to the transition $\alpha$. The last term in eq.~\ref{Hmatter} describes the dipole - dipole interaction between the intersubband plasmons, with a coupling constant:
\begin{equation} 
\hbar \Xi_{\alpha,\beta}= \frac{1}{2} C_{\alpha,\beta} \frac{E_{P_{\alpha}}E_{P_{\beta}}}{\sqrt{{\widetilde{E}_{\alpha}}\widetilde{E}_{\beta}}}.
\end{equation}
The quantity $C_{\alpha,\beta}$ is the plasmon-plasmon correlation coefficient~\cite{todorov_PRB2012} and it is very close to 1 in a square quantum well, but can be significantly different in a parabolic well ($\approx 0.7$). Note that the coupling constant $\Xi_{\alpha,\beta}$  depends on the electronic densities in the subbands and can be varied by adjusting the doping in the well. This coupling is a key feature to understand the collective response of the system, as it phase-locks the electronic oscillations, and yields new dressed excitations that can be obtained from the diagonalisation of $H_{\mathrm{matter}}$ (Eq.~\ref{Hmatter}). We indicate the energies of these dressed states $\hbar \Omega_J$, and $P_J^\dagger$ the corresponding bosonic creation operator. 
One can now proceed to the last step of the diagonalisation that deals with the interaction of the light with the new collective modes of the system. The Hamiltonian of the system can be now written in the form: 
\begin{equation}
\label{Htot}
H=E_c \left( a^{\dagger} a + \frac{1}{2}\right)+ 
\sum_J {\hbar \Omega_J P_J^\dagger P_J} +
i \sum_J{ \frac{\hbar \Omega_{P_J}}{2} \sqrt{F_w \frac{\omega_c}{\Omega_J}} \left( a^\dagger -a \right)\left( P_J ^\dagger + P_J \right)}
\end{equation}
Here $a^{\dagger}$ ($a$) are the creation (destruction) operators for a photon mode at energy $E_c=\hbar \omega_c$, $F_{w}$ is the overlap between the photon mode and the intersubband polarizations and the sums run over the new normal modes of the system. In equation \ref{Htot} the interaction between the light and each of the plasmon modes is weighted by an effective plasma frequency $\Omega_{P_J}$ that determines the amplitude of the absorption peaks. Each of the $\Omega_{P_J}$ can be expressed in terms of the plasma frequencies associated to the intersubband transitions and allows one to introduce an effective dielectric function of the system in the form:
\begin{equation}
\label{epsilon}
\frac{\epsilon}{\epsilon_{eff}\left( \omega \right)}=1+ \sum_J{\frac{{\Omega_{P_J}}^2}{\omega^2-{\Omega_J}^2+i\omega \gamma_J}}
\end{equation}
where $\gamma_J$ are the linewidths of the plasmon modes. The quantum description gives us a meaningful physical interpretation of the effective dielectric function in terms of coupled dipoles, originating from the individual intersubband transitions. 

\begin{figure}[ht]
\centering
\includegraphics[width=0.7\columnwidth]{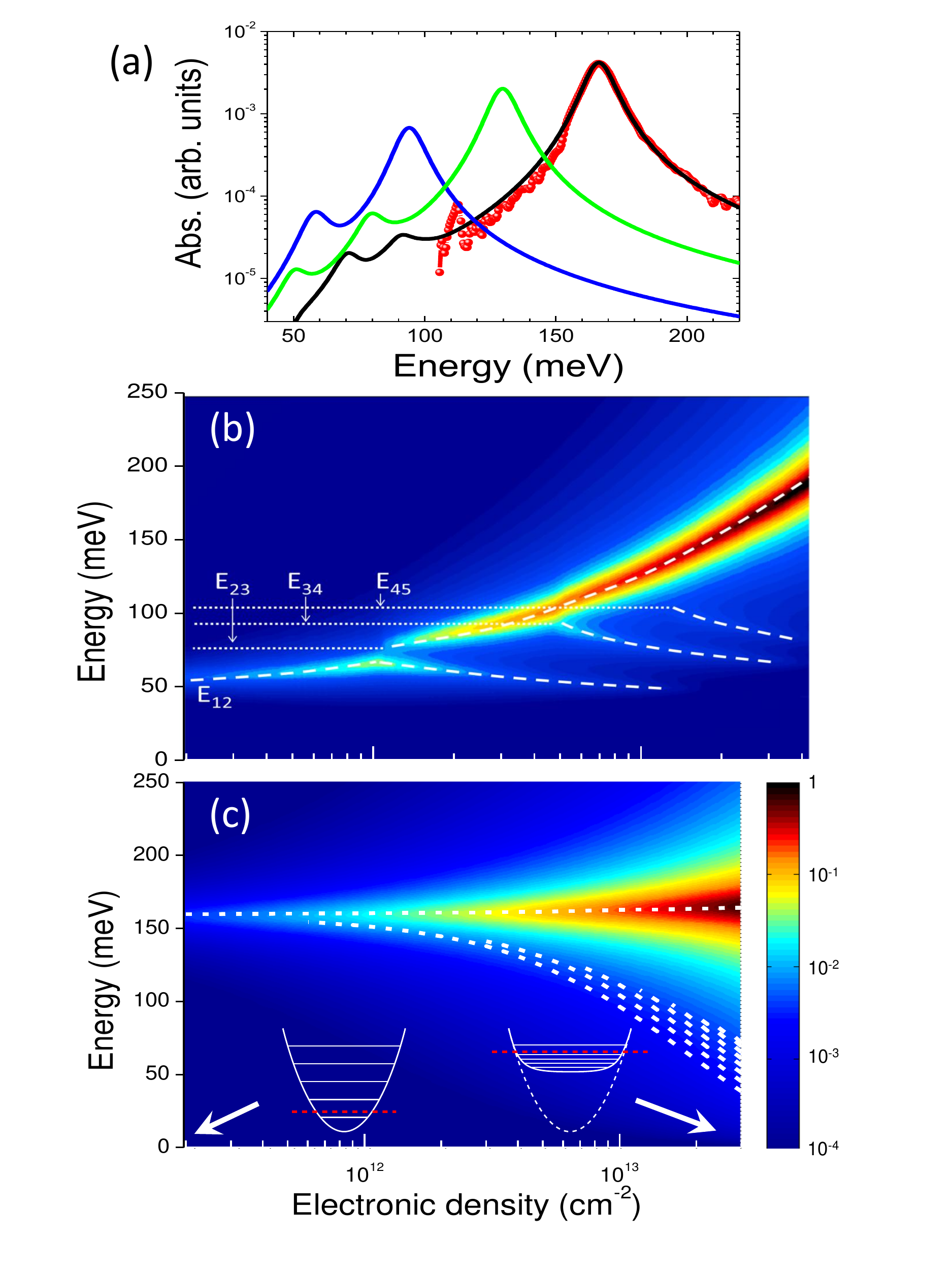}
\caption{(a) Simulated absorption spectra (at 77 K) in semi-logarithmic scale of a 18.5 nm GaInAs/AlInAs quantum well with 2 ($N_s=2.3 \times 10^{12}$ cm$^{-2}$, blue line), 3 ($N_s=9.2 \times 10^{12}$ cm$^{-2}$, green line) or 4 ($N_s=2.2 \times 10^{13}$ cm$^{-2}$, black line) occupied subbands. The red symbols present the measured absorption spectrum of the sample with $N_s=2.2 \times 10^{13}$ cm$^{-2}$ at 77K. (b) and (c): Simulated absorption spectra (logarithmic, color scale) as a function of the energy and electron sheet density for a square (18.5 nm thick) (b) and a parabolic (c) quantum well. The maximum value of the absorption in the graph has been normalized to one. The dashed lines indicate the dark and bright plasmons. In (c) we also sketch the parabolic potential shape and the electronic levels for the lowest and highest electronic densities used in the calculation. The Fermi energy is indicated with a (red) dashed line.}
\label{fig:multisubband}
\end{figure}

In order to account for all the experimental features and in particular non-parabolicity, we have implemented a semi-classical dielectric function to calculate the optical spectra. For this we have generalized Ando model~\cite{ando} to the case of several non-parabolic occupied subbands, in a similar way to~\cite{warburton_PRB}. Figure~\ref{fig:multisubband}a presents the calculated (at 77K) optical spectra when 2 ($N_s=2.3 \times 10^{12}$ cm$^{-2}$, blue line), 3 ($N_s=9.2 \times 10^{12}$ cm$^{-2}$, green line) or 4 ($N_s=2.2 \times 10^{13}$ cm$^{-2}$, black line) subbands are occupied. The spectrum obtained in the case of two occupied subbands is characterized by a renormalization of the energies of the two intersubband transitions and a redistribution of the amplitudes, as already observed by Warburton et al.~\cite{warburton_PRL}. As shown by the spectra in Fig.~\ref{fig:multisubband}a, when three or more subbands are occupied, the absorption amplitude is concentrated into a single peak, corresponding to the high energy (normal) mode of the coupled plasmon system, while the other absorption resonances have a negligeable amplitude. For the highest doping, corresponding to that of our sample, there are more than two decades between the high energy peak and the others. This is the regime of interest for our study. Indeed, when the electronic density is sufficiently high, the Coulomb dipole - dipole interaction induces a coherence in the system, and all dipoles oscillate in phase in the high energy mode. As a consequence, it concentrates the whole oscillator strength of the system. This is analogous to the center of mass motion of a chain of coupled mechanical oscillators that oscillate in phase.  We define {\textbf{bright multisubband plasmon}} this collective excitation of the 2DEG. As it concentrates the whole oscillator strength of the system, its effective plasma energy can be calculated as $\hbar \left( \Omega_P \right)_{\mathrm{bright}}=\sqrt{\sum _\alpha {E_{P_{\alpha}}^2}}$, where $\alpha$ labels the intersubband transitions. From this calculation, we find a value $\hbar \left( \Omega_P \right)_{\mathrm{bright}}=$136 meV. Note that the fact that $\hbar \left( \Omega_P \right)_{\mathrm{bright}}$ is comparable or greater than the energies of the bare intersubband transitions indicates the achievement of an unprecedented regime, in which the entire system responds with a single frequency to the optical excitation, despite the presence of several intersubband excitations at different frequencies. The red symbols in Fig.~\ref{fig:multisubband}a show the absorption spectrum measured at 77K. The agreement between the simulated and the experimental spectrum is excellent, for more than two decades~\cite{nota3}. 

The evolution of the multisubband plasmon with the doping in the well is illustrated in figure~\ref{fig:multisubband}b, where the calculated absorption (color logarithmic scale) is presented as a function of the energy and of the electron sheet density. The total absorption of the 2DEG increases monotonously with the electronic density, and concentrates into a single resonance, soon after $E_F$ is above the second subband. The bright mode shifts towards higher energies, while the normal modes displaying lower amplitudes (dark multisubband plasmons) have an opposite behavior: they red-shift with the electron density and their amplitude rapidly vanishes. This explains why in our experimental system we observe a single absorption peak. 

Our theoretical model is of general validity and can be applied to any potential shape.  In particular we verify that the Kohn theorem~\cite{kohn} is satisfied for a parabolic potential, by demonstrating that charge effects induced by Coulomb potential do not influence the absorption energy. This is shown in figure~\ref{fig:multisubband}c where the simulated absorption at 77K (color scale) for a parabolic quantum well is shown as a function of the energy and electronic sheet density~\cite{nota4}. The static charge in the quantized 2D-state of the parabolic well tends to flatten the potential (as shown in the figure), while the ensemble of the coupled oscillating dipoles, associated with different transitions, generates a sole bright multisubband plasmon. In a way, one could say that the plasma energy of the whole system compensates the reduced confinement due to the change in the potential shape. The energy evolution of the dark plasmons, as a function of the electron sheet density, is also indicated in the figure. It is worth noticing that, in the limit of very high charge density, the subband separation tends to zero and the absorption resonance corresponds to that of a 3D plasmon of a charged slab. This is an important difference from the case of a square well where the two dimensional character is always preserved. 

\begin{figure}[ht]
\centering
\includegraphics[width=0.7\columnwidth]{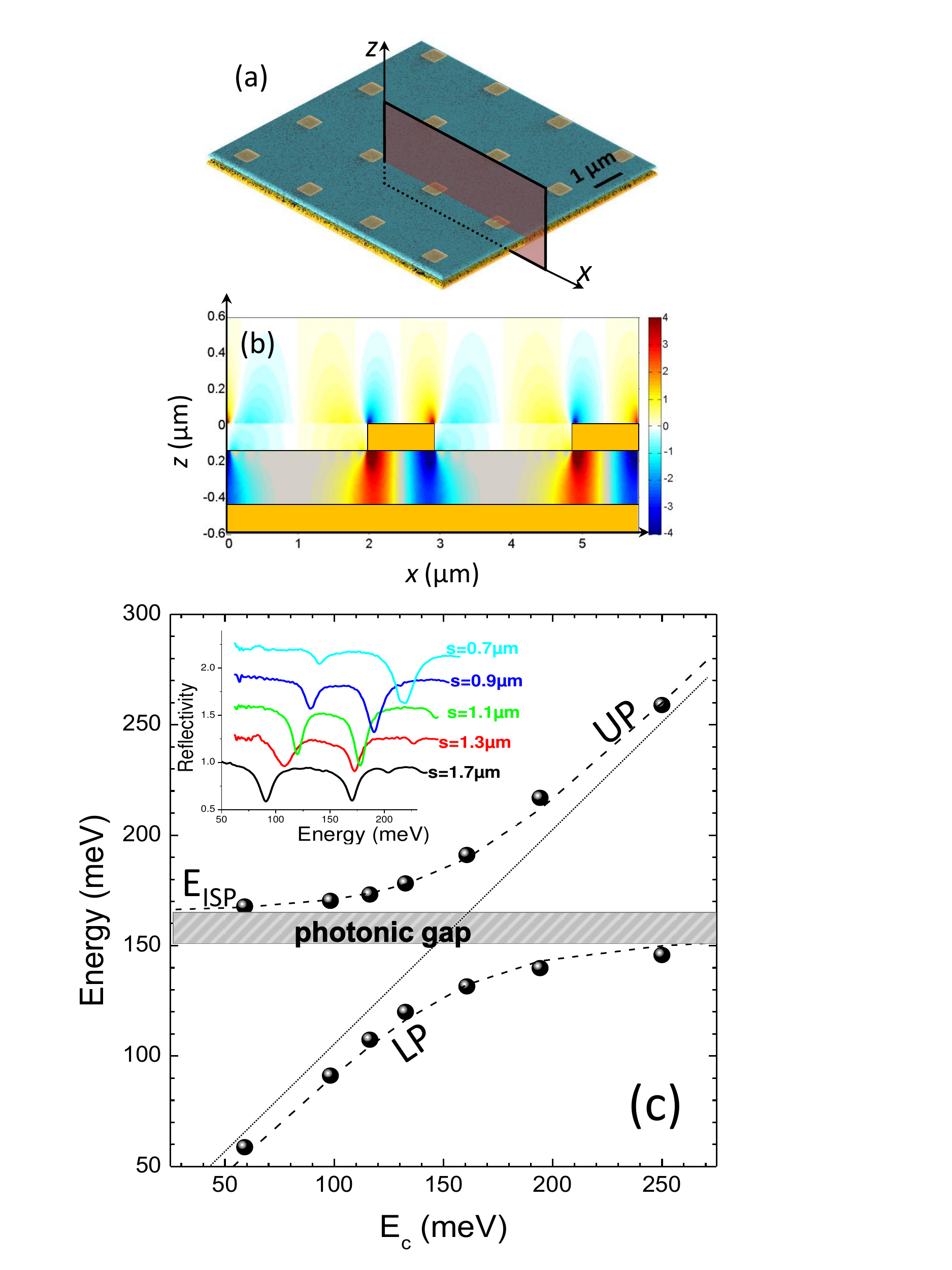}
\caption{(a) Scanning electron microscope image of the sample. The cavity thickness is 252 nm.  (b) $z$-component of the electric field for the fundamental mode, in the plane sketched in panel (a). (c) Symbols: Energy  position of the reflectivity minima measured at 300 K as a function of the energy of the cavity mode. The dashed lines indicate the polariton dispersion simulated by using eq.~\ref{eigen}. The two horizontal lines demarcate the photonic gap. In the inset we present some reflectivity spectra measured at 300 K for different values of the patch width. The cavity is filled with 5 GaInAs/AlInAs quantum wells (18.5 nm/30 nm).}
\label{fig:ultrastrong}
\end{figure}

The giant dipole moment of the multisubband plasmon can be directly observed by inserting the quantum well in a microcavity and measuring the ultra-strong coupling that arises from the interaction with the cavity mode~\cite{ciuti, gunter}. In that case the light-matter interaction gives rise to two intersubband polaritons, whose energies are obtained by diagonalizing the Hamiltonian in eq.~\ref{Htot}. In the cooperative regime, this procedure results in the following eigenvalue equation, describing the polariton dispersion: 
\begin{equation}
\label{eigen}
\left( E^2 - E_{MSP}^2 \right) \left( E^2-E_c^2\right)=E_R^2 E_c^2
\end{equation}
where $E_{MSP}=166$ meV is multisubband plasmon energy and $E_R=\hbar {\left(\Omega_P\right)}_{\mathrm{bright}} \sqrt{F_w}$ is the Rabi energy. This equation is analogous to that describing the ultra-strong coupling regime in a quantum well with only one occupied subband~\cite{todorov_PRL2010}. To investigate this regime, we have realized metal - dielectric - metal cavities~\cite{todorov_PRL2009}, in which light is confined in a highly subwavelength semiconductor layer \cite{jouy_APL}. Our sample contains 5 GaInAs/AlInAs quantum wells identical to the previous one, corresponding to a cavity thickness of only 252 nm and to a confinement factor $F_w= 0.17$. The top gold layer is patterned into a grating, composed by square patches, of width $s$ ($0.2 \, \mu$m $ < s < 5\,  \mu$m), separated by $2 \, \mu$m distance. Fig.~\ref{fig:ultrastrong}a presents a scanning electron microscope top view of the sample, with $s$ = 500 nm. The cavity mode energy depends on $1/s$, while the periodicity of the grating determines the coupling of the free space radiation with the cavity modes~\cite{jouy_APLplasmon, todorov_opex}. The $z$ component of the electric field of the fundamental mode is plotted in figure \ref{fig:ultrastrong}b, showing that light is confined under the gold patches. Reflectivity spectra measured at room temperature at an incidence angle of $10^\circ$ for different values of $s$ are shown in the inset of fig.~\ref{fig:ultrastrong}c. Two minima, associated to the upper and lower polaritons are visible. The polariton energies are plotted in the main panel of fig. \ref{fig:ultrastrong}c as a function of the cavity mode energy. The anticrossing in the dispersion allows measuring a Rabi energy $E_R = 57$ meV, corresponding to an effective plasma energy of $\hbar {\left(\Omega_P\right)}_{\mathrm{bright}} = 138$ meV, in agreement with the calculated value. The dashed lines indicate the polaritonic dispersion given by eq.~\ref{eigen}. Note that at 77 K we found $E_R = 58$ meV. This indicates that the coupling with light is very stable with respect to the temperature, in spite of the redistribution of the electrons within several subbands, similarly to a parabolic well~\cite{geiser}. This is very different from the situation of a quantum well with a single occupied subband. Indeed in this case the Rabi energy decreases between $ 10\% - 20\%$ when going from 77 K to 300 K \cite{jouy_APL}. The ratio between the Rabi energy and the plasmon resonance is 0.33, almost twice the best values obtained at room temperature in the mid infrared ~\cite{jouy_APL, anappara}. This result has been obtained thanks to the cooperative effect that allows us to increase at will the density of electrons interacting with the cavity mode. 

In conclusion, we demonstrated that, when several subbands of high density 2DEG are occupied, the system enters a many-body cooperative regime, in which the optical absorption is a single sharp resonance, associated to the excitation of a multisubband plasmon. When this collective resonance is coupled to a cavity mode, the system enters the ultra-strong coupling regime. The light-matter coupling induced by the Coulomb interaction between dipolar oscillators is reminiscent of a Dicke superradiant state~\cite{dicke} and opens the possibility to new coherent phenomena between the electronic states that can be controlled by adjusting the charge density. Finally, the giant collective dipole can be used for designing superradiant mid and far infrared emitters.

\begin{acknowledgments}
We thank P. Petit, M. L. Della Rocca, P. Lafarge and U. Gennser for help with Shubnikov - de Haas measurements. This work has been partially supported by the French National Research Agency (ANR) in the frame of its Nanotechnology and Nanosystems program P2N, Project No. ANR-09-NANO-007. We acknowledge financial support from the ERC grant "ADEQUATE".
\end{acknowledgments}

\newpage

\begin{thebibliography}{32}
\bibitem{nota1} Transitions between states with higher energy separation ($E_n \rightarrow E_{n+3}$) are also possible, however their oscillator strength never exceeds $\approx 3\%$ of those between adjacent states ($E_n \rightarrow E_{n+1}$) and are therefore neglected.
\bibitem{nota2} Note that, due to the strong non-parabolicity of our system, the energies of the absorption peaks are lower than the transition energies at $k_{//} = 0$.
\bibitem{nota3} We also obtain an excellent agreement between experimental and simulated absorption spectra at 300 K by taking into account the temperature dependence of the energy gap and of the electron distribution.
\bibitem{nota4} The barrier potential has been artificially set to 3.6 eV, in order to obtain a plasmon energy and an electronic density comparable to that of our square quantum well. If the conduction band discontinuity of GaInAs/AlInAs is considered, the parabolic potential is completely screened for an electronic density of $5.8 \times 10^{12}$ cm$^{-2}$.
\end{thebibliography}
\end{document}